**Response:"Kupczynski's Contextual Locally Causal Probabilistic Models are constrained by Bell's theorem"**

**Marian Kupczynski**

Department of computer science and engineering, University of Quebec in Outaouais (UQO), Case postale 1250, succursale Hull, Gatineau. QC, Canada J8X 3X7

Correspondence: marian.kupczynski@uqo.ca

**Abstract:** In our contextual model, statistical independence is violated, thus it is not constrained by Bell's Theorem. Individual outcomes are created locally in a deterministic way in a function of setting dependent variables describing measuring instruments and variables describing physical systems, at the moment of their interactions. These setting dependent variables may be correlated, but not necessarily due to spooky influences or super-determinism. In several Bell Tests, two time series of distant clicks are converted into finite samples containing pairs of non-zero outcomes. This data not only violates Bell –CHSH inequalities but also no-signalling. Our model allows explaining both the raw and the final data in these experiments. Moreover, our model, does not compromise experimenters' freedom of choice. The violation of no-signalling is neither consistent with quantum description of an ideal EPRB experiment nor with a standard local realistic and stochastic hidden variable models, thus the title of this article and its conclusion:" Kupczynski's escape route for local realism is not available" are misleading.

1    **Introduction**

In [1], Richard Gill and Justo Lambare (G-L), using similar arguments as in [2, 3], criticize our paper [4] and claim that their criticism apply to the whole sequence of our papers [5-8].  We replied to [2] in [9] and to [3] in [10], explaining that our papers do not contain false statements and that our models are not a special case of a local hidden variable model (LHVM) [11, 12].  Apparently our explanation was not clear enough, because G-L added two new sections and continue insisting that our model is constrained by Bell's Theorem.

In this short note we are not going to reproduce all the arguments given in [9, 10]. After resuming shortly, why we disagree with G-L, we discuss, in some detail, the content of two new sections added in [1].

In LHVM,  "entangled pairs" are described as they were pairs of socks or pairs of dice. It seems, that G-L agree with it.  In quantum phenomena measuring instruments play an active role and measurement outcomes depend strongly on experimental contexts [8, 13-16]. Therefore the assumption *of statistical independence* (SD), according to which random variables in LHVM do not depend on experimental contexts, is an unrealistic simplification [16-19]. Already in 1964, Bell clearly pointed out that, if SD was violated [11], then his inequalities could not be derived.

The data in several Bell Tests e.g. [20] not only violate Bell –CHSH inequalities, but also no-signalling, what can nether be explained by quantum mechanics (QM) nor by LHVM. This is why, in our contextual hidden variable model (CHVM) we incorporate setting dependent instrument variables, which together with variables describing physical systems locally determine the values of experimental outcomes (a, b) in each experimental setting (x, y). In fact we have 4 probabilistic models each defined on 4 disjoint specific probabilistic spaces $\Lambda_{xy}$ .

Using these models we can describe 4 sets of raw and final experimental data. For each setting , using synchronized time windows of width W, two distant time series of clicks are converted into raw data containing pairs of outcomes (a, b), where a, b=0, +/-1. Next by post-selecting only non- zero outcomes the final data are obtained and used to estimate pair-wise expectations and test various inequalities and no-signalling [6,8].

In our approach, the raw data are described by a probabilistic model M1 and the final data by M2. In M1, there are no a priori constraints on pairwise expectations and M1 explicitly violates SD. Moreover, probabilistic models describing different settings use disjoint probability spaces, thus in agreement with [21, 22] we correctly concluded that CHSH may not be derived. Probably, we should have said, that they cannot be directly derived, because G-L constructed a counterfactual probabilistic model, which apparently allows to prove CHSH for M1 in the sense of a probabilistic coupling [1-3]. The existence of a probabilistic coupling does not mean that the CHSH inequalities hold in the experiments it tries to describe. It means that, if experiment follows a specific experimental protocol inspired by this coupling, then there exist probabilistic bounds [23] on a possible violation of these inequalities, in this case, by the raw data. In Bell Tests, nobody is checking CHSH inequalities using the raw data. The model M1 is used <u>only</u> as the first step from which a probabilistic model M2 describing the <u>final data is derived.</u> A <u>probabilistic coupling does not exist</u> for M2 and Bell-CHSH inequalities cannot be derived.

Therefore, the existence or nonexistence of G-L coupling for the raw data is irrelevant for the soundness of reasoning and conclusions of a sequence of papers [4-8]. In our model outcomes are produced in a deterministic and local way, but since the correlations and experimental context are nonlocal, thus our model can neither be called local nor non- local. Nevertheless, the violation of SD in M2 is neither due to spooky influences nor to super-determinism. Therefore, the result of these experiments allows rejecting *local realism* understood as *predetermination of experimental outcomes* but do not justify the speculations about the quantum magic.

Recent Bell Tests e.g. [24] cannot be described by M2, but they can be described by a more general CHVM, which we call in this paper M3. In M3, the SD is also violated but not necessarily by spooky influences.

In section 2, we discuss probabilistic models M2 and M3 mentioned above.

In section 3, we analyze some new arguments given by G-L in [1].

Section 4 contains some conclusions.

.
## 2    Details of our contextual hidden variable models

We describe Bell Tests, which uses an experimental protocol discussed in the introduction.

- Photonic signals are described by variables $(\lambda_1, \lambda_2) \in \Lambda_1 \times \Lambda_2$ and $p(\lambda_1, \lambda_2)$.
- In a setting (x, y), Alice's and Bob's instruments, at the moment of the measurement, are described by variables $(\lambda_x, \lambda_y) \in \Lambda_x \times \Lambda_y$ and probability distributions $p_x(\lambda_x)$ and $p_y(\lambda_y)$.
- Outcomes $0, \pm 1$ are the values of setting dependent functions $A_x(\lambda_1, \lambda_x)$ and $B_y(\lambda_2, \lambda_y) = 0, \pm 1$.
- Expectation values of inconsistently connected random variables $A_{xy}$ and $B_{xy}$, describing the final data are given by:

$$E(A_{xy} B_{xy}) = E(A_x B_y | A_x B_y \neq 0) = \sum_{\lambda \in \Lambda'_{xy}} A_x(\lambda_1, \lambda_x) B_y(\lambda_2, \lambda_y) p_{xy}(\lambda) / C_{xy} \qquad (1)$$

$$E(A_{xy}) = E(A_x | A_x B_y \neq 0) = \sum_{\lambda \in \Lambda'_{xy}} A_x(\lambda_1, \lambda_x) p_{xy}(\lambda) / C_{xy} \qquad (2)$$

$$E(B_{xy}) = E(B_y | A_x B_y \neq 0) = \sum_{\lambda \in \Lambda'_{xy}} B_y(\lambda_2, \lambda_y) p_{xy}(\lambda) / C_{xy} \qquad (3)$$

where $C_{xy} = p(A_x B_y \neq 0)$, $p_{xy}(\lambda) = p_x(\lambda_x) p_y(\lambda_y) p(\lambda_1, \lambda_2)$, $\Lambda_{xy} = \Lambda_1 \times \Lambda_2 \times \Lambda_x \times \Lambda_y$ and $\Lambda'_{xy} = \{\lambda \in \Lambda_{xy} | A_x(\lambda_1, \lambda_x) \neq 0, B_y(\lambda_2, \lambda_y) \neq 0\}$.

In the papers [4-8], the constants $C_{xy}$, were missing but it does not change any conclusions because the equations (1-3), were never used to make quantitative predictions.

The equations (1-3) define our model M2, which, as we explained in [8, 13] is a special case of the model M3 defined below :

$$E(A_{xy} B_{xy}) = E(A_x B_y | x, y) = \sum_{\lambda \in \Lambda'_{xy}} A_x(\lambda_1, \lambda_x) B_y(\lambda_2, \lambda_y) p_{xy}(\lambda_x, \lambda_y) p(\lambda_1, \lambda_2) \qquad (4)$$

or more generally:

$$E(A_{xy} B_{xy}) = \int_{\Lambda_{xy}} A_x(\lambda_1, \lambda_x) B_y(\lambda_2, \lambda_y) \rho_{xy}(\lambda_x, \lambda_y) \rho(\lambda_1, \lambda_2) d\lambda_x d\lambda_y d\lambda_1 d\lambda_2 \qquad (5)$$

In M3, we have in fact 8 random variables instead of 4 because, as it was explained in the Contextuality-by- Default approach the random variables $A_{xy}$ and $A_{xy'}$ a priori stochastically unrelated [8, 25-27]. In M3, instrument variables are explicitly correlated, this dependence is more likely related to global space-time symmetries than to spooky influences. Moreover, M2 and M3 violate SD without restricting *experimenters' freedom of choice* [5, 8,13]

## 3    Reply to specific sections in the paper [1]

**In section 2** : G-L claim incorrectly, that our statement in the abstract of the paper [4]:

> *Since variables (A, A') and (B, B') cannot be measured jointly, neither N × 4 spread- sheets nor a joint probability distribution of (A, A', B, B') exist…* is false and fallacious. They misinterpret Fine's Theorem [28]. It is well known, that a joint probability distribution (JP) of *(A, A', B, B')* describes a random experiment in which, in each trial, 4 outcomes are outputted and displayed in Nx4 spreadsheet. Only if JP exists, one may derive rigorously Bell-CHSH inequalities for some cyclic combinations of pairwise expectations [29]. These expectations, estimated using the same Nx4 experimental spreadsheet, can never violate these inequalities. We can say then, that Bell-CHSH inequalities hold in this experiment.

JP and Nx4 spreadsheets do not exist in an ideal EPRB experiment, in Bell Tests and in Lambare- Franco (L-F) counter-example [13, 30]. Pairwise expectations, estimated using 4 different Mx2 experimental spreadsheets, may easily violate Bell-CHSH inequalities. Nevertheless, one may postulate the existence of a probabilistic coupling, motivated by some physical/metaphysical assumptions (e.g. local realism/counterfactual definiteness), and test its plausibility. This is what was tested in various spin polarization correlation experiments. For Bell local realistic model and for L–F counter-example JP's exist, because the outcomes are predetermined, but only in the sense of a probabilistic coupling.

In [13], we discussed this issue with all necessary details for Bell local realistic model but , because of some misprints, our discussion of L-F counter-example was perhaps, less clear. Therefore, we explain it again below.

We have: 7 random variables: L taking values $\lambda \in \Lambda = \{1, 2, 3, 4, 5, 6\}$, *X* taking values x={1.-1}, *Y* taking values y={1.-1} , $A_x$ and $B_y$. L describes an experiment in which hidden variables are sampled from $\Lambda$ by rolling a dice, X and Y are random variables describing flipping fair coins in order to determine experimental settings (x, y), $A_x$ and $B_y$ are random variables describing a statistical scatter of experimental outcomes : $a = x^\lambda$ and ; $b = y^{\lambda+1}$ .

There are 4 incompatible experiments, labelled by (x, y), and only 2 outcomes are outputted in each trial, thus JP of 4 random variables ($A_1$, $A_{-1}$, $B_1$, $B_{-1}$) does not exist and estimates of pairwise expectations $E(A_x B_y)$ may violate CHSH inequalities.

For each pair of settings (x, y) outcomes are predetermined by the values of $\lambda$, thus there exists a set of 4 jointly distributed random variables ($A'_1$, $A'_{-1}$, $B'_1$, $B'_{-1}$), defining a probabilistic coupling : $E(A_x)=E(A'_x)$, $E(B_y) = E(B'_y)$, ), $E(A_x B_y)= E(A'_x B'_y)$.

Namely:

$$A'_x = A'_x(L) = x^L; \quad B'_y = B'_y(L) = y^{L+1} . \qquad (6)$$

Variables ($A'_1$, $A'_{-1}$, $B'_1$, $B'_{-1}$) define a mapping $M : \Lambda \Rightarrow \Omega = \{(1,1,1,-1),(1,-1,1,1)\}$ and describe a different random experiment in which in each trial one of two quadruplets is outputted with probability ½. Marginal pairwise expectations are equal: $E(A'_1 B'_1) = 1, E(A'_1 B'_{-1}) = E(A'_{-1} B'_1) = 0, E(A'_{-1} B'_{-1}) = -1$ and their cyclic combinations strictly obey CHSH inequalities. Experimental outcomes can be displayed in an Nx4 spreadsheet and this spreadsheet can be used to estimate pairwise expectations. Then, CHSH inequalities hold for any N.

The existence of the probabilistic coupling (1) does not mean that CHSH inequalities hold for 4 incompatible experiments we started with. The existence of probabilistic coupling allows only deriving probabilistic bounds on the significance of violation of CHSH in these experiments [13, 23].

In [24] L-F concluded: "*according to Fine's theorem A, a joint probability P ($A_1$; $A_{-1}$; $B_1$; $B_{-1}$) exists, although the experiments are incompatible*". We explained above, why this statement is imprecise. It is important to understand, that JP in Fine theorem may exists, in the context of EPRB experiments, but only in the sense of a probabilistic coupling.

One should keep in mind that there are always two sets of random variables: one set describes finite samples and a scatter of outcomes in particular experiments and another is a part of a probabilistic model used to describe random experiment(s) as a whole. An instructive example discussing correlations in the collisions of metal balls may be found in [4].

**In section 4**, G-L reproduce from [2, 3] their construction of a probabilistic coupling for a model M1 and they claim, without proof, that such construction is also possible for the model M2 and M3 , what is not true.

**In section 5**, G-L literary reproduce the content of the article [28]. In this section they use a different notation; the inputs are denoted (a, b) and the outputs (x, y). It is not important .Figures 1 and 2, have been known, since many years, and they describe so called Bell-game, which is an over-simplified model of real spin polarization correlation experiments. Outcomes of Bell games may be described by a local stochastic hidden variable model and Bell-CHSH inequalities may be derived using an appropriate probabilistic coupling. The DAG and the explanations given in this sections do not apply to our CHVM models (1-5) in which CHSH inequalities cannot be derived..

Therefore, the conclusion :

> $X = f(A, \Lambda)$, $Y = g(B, \Lambda)$, where $f$, $g$ are some functions and $(A, B)$ is statistically independent of $\Lambda$. This is what Bell called a local hidden variables model. It is absolutely clear that Kupczynski's notion of a probabilistic contextual local causal model is of this form.

is unfounded and misleading. Our detailed review of the article [31] was posted on Qeios [32]

In a recent paper [33], Richard Gill seems to agree that our CHVM model (5) is not a special case of LHVM and that it is not constrained by Bell Theorem. However, he dismisses this model, because using it one can explain any imperfect correlations between the outputs of distant experiments. We do not dismiss the Newton's equations because they do not make any predictions, if we do not specify specific forces and initial conditions. Using (5), and assuming specific statistical dependence between the instrument variables, we may hope to reproduce $\cos^2(\theta_x - \theta_y)$ dependence of of these correlations. The paper [33] contains other questionable statements and conclusions which we discuss in detail in our review [34], posted on Qeios.

In [4, 9, 32,34], we explained that G-L misrepresented the content and conclusions of several of our papers. In science one may disagree and argue using strictly scientific arguments. In our opinion, several statements in [1-3] were inappropriate.

G-L also incorrectly claimed, that their criticism applies to the papers of other authors [], who also questioned the rationale of using the same probability space to describe 4 stochastically independent random experiments and extraordinary metaphysical speculations claimed to be justified by the results of Bell Tests.

## 3   Conclusions

Our contextual models are not constrained by Bell Theorem In our papers [4-8, 13], we explain that evoking quantum magic, when discussing the results of Bell Tests is unfounded and counterproductive.

Sound arguments against quantum nonlocality and in favor of contextual probabilistic models of quantum phenomena may be found in an excellent Khrennikov's review article [35].

Convincing and detailed arguments, that the metaphysical implications of recent Bell Tests are quite limited have been given recently and independently by Dzhafarov [36] and by DeRaedt et al. [37]

The inequalities are violated in physics and in cognitive science, but it neither proves the completeness of quantum mechanics [5, 1, 16, 40] nor nonlocality of nature [4, 35, 41-47].

### References

1. Gill R. D. and Lambare J.P.(2023), Kupczynski's Contextual Locally Causal Probabilistic Models are constrained by Bell's theorem, arXiv:2208.09930**v6** [quant-ph]; https://doi.org/10.48550/arXiv.2208.09930
2. Gill R. D. and Lambare J.P.(2022), Kupczynski's contextual setting-dependent parameters offer no escape from Bell-CHSH, arXiv:2208.09930**v1** [quant-ph], https://doi.org/10.48550/arXiv.2208.09930
3. Gill RD and Lambare JP (2022), General commentary: Is the moon there if nobody looks—Bell inequalities and physical reality. Front. Phys. **10**:1024718 (3 pp.) doi:10.3389/fphy.2022.1024718 534
4. M. Kupczynski (2020) Is the moon there if nobody looks: Bell inequalities and physical reality.Frontiers in Physics **8** (13 pp.) htps://www.frontiersin.org/articles/10.3389/fphy.2020.00273/full 53
5. Kupczynski, M., (2017) Can we close the Bohr–Einstein quantum debate? Phil. Trans. R. Soc. A 375 20160392. http://dx.doi.org/10.1098/rsta.2016.0392
6. Kupczynski, M. (2017) Is Einsteinian no-signalling violated in Bell tests? Open Phys. 2017 5 739–753. https://www.degruyter.com/document/doi/10.1515/phys-2017-0087
7. Kupczynski,M. (2018) Quantum mechanics and modelling of physical reality. Physica Scripta 93 123001. https://iopscience.iop.org/article/10.1088/1402-4896/aae212, https://arxiv.org/abs/1804.02288
8. Kupczynski, M (2021) Contextuality-by-default description of Bell tests: Contextuality as the rule and not as an exception. Entropy **23**(9), 1104 (15pp.) https://doi.org/10.3390/e23091104
9. Kupczynski M, (2022) Is the Moon there if nobody looks: A reply to Gill and Lambare, arXiv:2209.07992 [quant-ph] http://arxiv.org/abs/2209.07992
10. Kupczynski M. (2023), Response: "Commentary: Is the moon there if nobody looks? Bell inequalities and physical reality". Front. Phys. **11**:1117843 (3pp). doi:10.3389/fphy.2023.1117843
11. Bell, J.S. (1964) On the Einstein-Podolsky-Rosen paradox.Physics1964, 1, 195-200.
12. Bell, J. S., Speakable and Unspeakable in Quantum Mechanics. Cambridge UP, Cambridge (2004)

13. Kupczynski, M. (2023) Contextuality or Nonlocality: What Would John Bell Choose Today? Entropy **25**(2), 280 (13 pp.) https://doi.org/10.3390/e25020280
14. Bohr, N. The Philosophical Writings of Niels Bohr. Ox Bow Press: Woodbridge, UK, 1987.
15. Khrennikov, A. Yu.Contextual Approach to Quantum Formalism . Springer, Dortrecht (2009)
16. Kupczynski, M.(2006). Seventy years of the EPR paradox. *AIP Conf. Proc*.**2006**,861, 516-523
17. Nieuwenhuizen T.M., Where Bell went wrong, AIP Conf. Proc., 2009, 1101, 127-33.
18. Nieuwenhuizen T.M., Is the contextuality loophole fatal for the derivation of Bell inequalities, Found. Phys. 2011, 41, 580-591.
19. Nieuwenhuizen, T.M.; Kupczynski, M. The contextuality loophole is fatal for derivation of bell inequalities: reply to a comment by I. Schmelzer.Found Phys.(2017)47:316–9. doi: 10.1007/s10701-017-0062-y
20. Weihs G.; Jennewein T.; Simon C.; Weinfurther H.; Zeilinger A. Violation of Bell's inequality under strict Einstein locality conditions*, Phys. Rev. Lett.* , **1998**, 81, 5039-5043.
21. Larsson, J., (1998), Bell's inequality and detector inefficiency, *Physical Review A. Atomic, Molecular,and Optical Physics*, 57(5), 3304-3308. https://doi.org/10.1103/PhysRevA.57.3304
22. Larsson J.-A° , Gill R. D. , Bells inequality and the coincidence time loophole, Europhys. Lett. 67, 707-713 (2004)
23. Gill, R.D. (2014), Statistics, Causality and Bells Theorem, Statist. Sci. 29(4): 512-528. https:doi.org/10.1214/14-STS490
24. Rosenfeld, W. et al. Event-Ready Bell Test Using Entangled Atoms Simultaneously Closing Detection and Locality Loopholes, Phys. Rev. Lett. 119, 010402 (2017)
25. Dzhafarov, E.N. and Kujala J.V.(2014) Contextuality is about identity of random variables. Physica Scripta 2014, T163:014009.
26. Dzhafarov E.N., Kujala J.V. and Larsson J.-Å. (2015)Contextuality in three types of quantum-mechanical systems. Foundations of Physics 2015, 7, 762-782.
27. Kujala J.V.,Dzhafarov E.N. and Larsson J-Å, (2015),Necessary and sufficient conditions for extended noncontextuality in a broad class of quantum mechanical systems. Physical Review Letters 2015, 115:150401
28. Fine, A.(1982) Hidden variables, joint probability and the Bell inequalities, Phys. Rev. Lett., 1982, 48, 291-295.
29. Araujo M.;Quintino M.T.;Budroni C.;Cunha M.T.; Cabello A.(2013), All noncontextuality inequalities for the n-cycle scenario. Phys. Rev. A2013, 88, 022118 ;https://arxiv.org/pdf/1206.3212.pdf
30. Lambare J.P. and Franco R.(2021) , A Note on Bell's Theorem Logical Consistency. *Found Phys* **51,** 84 (2021). https://doi.org/10.1007/s10701-021-00488-z.
31. Gill R. D,(2023) Bell's theorem is an exercise in the statistical theory of causality, arXiv:2211.05569v2 [quant-ph**]** for this version); https://doi.org/10.48550/arXiv.2211.05569

32. Kupczynski, M. (2023), Review of: 'Bell's theorem is an exercise in the statistical theory of causality', https://www.qeios.com/read/DRHFO9
33. Gill R.D. (2023), Further comments on 'Is the moon there if nobody looks? Bell inequalities and physical reality', https://doi.org/10.32388/OLGL3Z
34. Kupczynski, M. (2023), Review of: 'Further comments on 'Is the moon there if nobody looks? Bell inequalities and physical reality', https://doi.org/10.32388/J03OQN
35. Bell, J.S. (1964) On the Einstein-Podolsky-Rosen paradox. Physics 1964, 1, 195-200.
36. Bell, J. S., Speakable and Unspeakable in Quantum Mechanics. Cambridge UP, Cambridge (2004)
37. Khrennikov, A.(2022), Contextuality, Complementarity, Signaling, and Bell Tests, Entropy 2022, 24 (10), 1380; https://doi.org/10.3390/e24101380 - 28 Sep 2022
38. Dzhafarov E. N.(2021), Assumption-Free Derivation of the Bell-TypeCriteria of contextuality/Nonlocality, *Entropy* 2021, *23*(11),1543 **https://doi.org/10.3390/e23111543**43
39. De Raedt H, et al. (2023), Einstein-Podolsky-Rosen-Bohm experiments: a discrete data driven approach, Annals of Physics, Volume 453, 169314, 2023; https://doi.org/10.1016/j.aop.2023.169314
40. Kupczynski. M.(1987). Bertrand's paradox and Bell's inequalities. *Phys.Lett. A* 1987, 121,205-07
41. Kupczynski M.,(2018) Closing the Door on Quantum Nonlocality, Entropy, 2018, 20, https://doi.org/10.3390/e20110877
42. Khrennikov, A.(2019) Get rid of nonlocality from quantum physics. *Entropy* **2019**, *21*, 806.
43. Khrennikov, A.(2020) Two Faced Janus of Quantum Nonlocality, *Entropy* **2020**, *22*(3), 303; **https://doi.org/10.3390/e22030303**
44. Hess. K. ;Philipp, W. (2005) Bell's theorem: critique of proofs with and without inequalities. *AIP Conf. Proc.***2005**, 750, 150
45. Hess, K.(2022): A Critical Review of Works Pertinent to the Einstein-Bohr Debate and Bell's Theorem *Symmetry* **2022**, *14*(1), 163; **https://doi.org/10.3390/sym14010163**
46. Hance, J. R.; Hossenfelder, S. ; Palmer, T. N. Supermeasured: Violating Bell-Statistical Independence without violating physical statistical independence*Found. Phys*. **2022**, 52, 81 .
47. Hance J.R.; Hossenfelder S. (2022), Bell's theorem allows local theories of quantum mechanics, *Nature Physics*, https://doi.org/10.1038/s41567-022-01831-5